\documentstyle[12pt,aasms4,psfig]{article}
\received{}
\accepted{}

\slugcomment{To appear in ApJ}
\lefthead{Wilkes et al.}
\righthead{BeppoSAX Observations of the Maser Sy2 Galaxy: ESO103-G35.}
\def\etal   {{\it et~al.}}
\def\alpe {$\alpha_E$}
\def\nh {${\rm N_H}$}
\def\lax    {${_<\atop^{\sim}}$}

\begin{document}

\title{BeppoSAX Observations of the Maser Sy2 Galaxy: ESO103-G35.}
\author{Belinda J. Wilkes$^1$, Smita Mathur,$^{1, 2}$Fabrizio Fiore,$^{3,4}$ Angelo
Antonelli,$^{3, 4}$ \& Fabrizio Nicastro$^{1}$ 
}
\affil{1: Harvard-Smithsonian Center for Astrophysics, Cambridge, MA 02138}
\affil{2: The Ohio State University}
\affil{3: BeppoSAX Science Data Center, via Corcolle 19, I-00131, Roma, Italy}
\affil{4: Osservatorio Astronomico di Roma, via Frascati 33, I-00040
Monteporzio, Italy}

\begin{abstract}
We have made BeppoSAX observations of the Seyfert 2/1.9
galaxy ESO103-G35, which contains a nuclear maser source and is known
to be heavily absorbed in the X-rays.  Analysis
of the X-ray spectra observed by SAX  in October 1996 and 1997 yields
a spectral index $\alpha_E$ = 0.74$\pm$0.07 (F$_{\nu} \propto
\nu^{-\alpha_E}$) which is typical of Seyfert galaxies and consistent
with earlier observations of this source. 
The strong, soft X-ray absorption has a column density,
$N_H$ of 1.79$\pm0.09 \times 10^{23}$ cm$^{-2}$, again consistent with earlier 
results.
The best fitting spectrum is that of a power law with a high energy
cutoff at 29$\pm$10 keV, a cold (E=6.3$\pm$0.1 keV, rest frame),
marginally resolved ($\sigma$ = 0.35$\pm$0.14 keV,
FWHM $\sim 31 \pm 12 \times 10^3$ km s$^{-1}$)
Fe K$\alpha$ line with EW 290$^{+100}_{-80}$ eV (1996) and
a mildly ionized Fe K-edge at 7.37$^{+0.15}_{-0.21}$ keV, $\tau =
0.24^{+0.06}_{-0.09}$. 
The K$\alpha$ line and cold absorption are consistent with
origin in a accretion disk/torus through which our
line-of-sight passes at a radial distance of $\sim 0.01$ pc. The Fe K-edge
is mildly ionized suggesting the presence of ionized gas probably
in the inner accretion disk, close to the central source or in a
separate warm absorber.
The data quality is too low to distinguish between these possibilities
but the edge-on geometry implied by the water maser emission favors the
former.
Comparison with earlier observations of ESO103-G35
shows little/no change in spectral parameters while the flux changes by
factors of a few on timescales of a few months.
The 2--10 keV flux decreased by a factor of $\sim 2.7$ between Oct 1996
and Oct 1997 with no detectable change in
the count rate $>$20 keV (i.e. the PDS data). Spectral fits to the
combined 
datasets indicate either a significant hardening of the spectrum 
(\alpe$\sim$0.5)
or a $\sim$constant or delayed response reflection component.
The high energy cutoff
(29$\pm$10 keV) is lower than the typical $\sim 300$ keV values seen in
Seyfert galaxies. A significant subset of similar sources would affect
current models of the AGN contribution to the cosmic X-ray background (CXRB)
which generally assume a high energy cutoff of $\sim 300$ keV.
\end{abstract}


\section{Introduction}
The sub-parsec masing disk recently found to be orbiting a $\sim 10^7$
solar mass black hole in the Seyfert 2 galaxy, NGC 4258, provides the
most compelling evidence to date for the existence of a massive black
hole in the nucleus of a galaxy (Miyoshi \etal\ 1995).
The disk is edge-on, the X-ray spectrum is heavily absorbed
(Makishima \etal\ 1995) and an active nucleus is visible in
polarized light (Wilkes \etal\ 1995).
Nearly all AGNs which contain maser sources and have
X-ray observations show strong X-ray absorption consistent with an
edge-on disk, suggesting that they generally harbor active
nuclei in their cores.
This is consistent with the generally accepted scenario that many Seyfert 2
galaxies are edge-on Seyfert 1 galaxies (Antonucci \& Miller 1985).
The combined power of the high-resolution radio observations and the
X-ray spectrum to study the structure of the disk both along the
line-of-sight and in the plane of the sky makes these objects potentially
key to our understanding of the inner regions of active galaxies.

The Seyfert 1.9/2 galaxy ESO103-G35 is a strong X-ray source, originally
discovered by HEAO 1/A2 (H1834-653, Marshall \etal\ 1979,
Piccinotti \etal\ 1982).
It was identified as a Seyfert 1.9 galaxy at a redshift of 0.0133
by Phillips \etal\ (1979)
though later spectra have shown no evidence for broad lines 
(Morris \& Ward 1988) leaving its  classification in some doubt.
It has been observed extensively in the X-rays. EXOSAT observations
revealed strong, soft X-ray absorption (\nh $\sim 2 \times 10^{23}$ cm${-2}$)
which showed a factor of 2 variation 
over 90 days (Warwick, Pounds \& Turner 1988), providing some of the first
evidence for variable absorption in a Seyfert nucleus.
{\it Ginga} observations revealed an equivalent width $\sim 330$ eV
Fe K$\alpha$ emission line
whose origin was not clear given the presumed (but not confirmed) presence
of a flaring source within the {\it Ginga} beam
during a portion of this observation (Warwick \etal\ 1993).
The continuum spectrum was consistent with that reported by EXOSAT.

ASCA observations confirmed and extended the earlier studies, reporting
a cold, resolved Fe K$\alpha$ emission line and a mildly ionized Fe K-edge
(Turner \etal\ 1997, Forster, Leighly \& Kay 1999). While flux variations up to a
factor of a few are common on short timescales, no significant spectral or
absorption variations have been detected.

Recent radio measurements have revealed water maser emission in the form of
a single line at $+100$ km s$^{-1}$ (Braatz, Wilson \& Henkel 1997)
with respect to the AGN redshift. Due to the
large negative declination ($\delta \sim -65^o$) of this source, higher 
resolution and sensitivity observations to further study this maser are 
difficult at present. However the combination of water maser emission
and the strong, absorbed X-ray source suggest a strong parallel with NGC4258
and motivated our observations with BeppoSAX to obtain an X-ray spectrum 
simultaneously over a wide energy band (1--200 keV). The moderate
spatial resolution ($\sim 2'$) of BeppoSAX facilitates identification of
any nearby,
contaminating sources and so avoids the confusion present in the {\it
Ginga} data. Here we present an analysis of the BeppoSAX X-ray data.

\section{Observations and Analysis.}
ESO103-G35 was observed with BeppoSAX at two epochs: Oct 1996 and Oct 1997
(see Table~\ref{tbobs}). Data were obtained with both gas scintillator
proportional counter (GSPC) detectors: low and medium energy
concentrator spectrometers (LECS and MECS respectively, Parmar \etal\
1990) and with the Phoswich detector system (PDS, Frontera \etal\
1991). The source was clearly detected and unresolved in all three
detectors and no nearby source of comparable strength was seen.

Counts were extracted from a $4'$ circle in both LECS and MECS detectors
and background counts subtracted, estimated
from the background file appropriate for 
this extraction radius provided by the BeppoSAX Data Center (SDC) as of March 1998.
The resulting net source count rates for the energy ranges over which the
instrument calibration is reliable are listed in 
Table~\ref{tbobs}. The PDS data were background subtracted as part of the 
data reduction pipeline.

\begin{table}[h]
\caption{Observational details for the BeppoSAX observation of ESO103-G35.}
\label{tbobs}
\begin{tabular}{llllll}
\hline
Start Date & End Date & Instrument & Exp. Time & Energies & Count Rate \\
&&& (ksecs) & (keV) &  (cts s$^{-1}$)  \\
\hline
3 Oct 1996 & 4 Oct 1996 & LECS & 10.238 & 0.1--5.0 & 0.031$\pm$0.002 
\\
&& MECS & 50.615 & 1.7--10.5 & 0.306$\pm$0.003 \\
&& PDS & 21.029 & 15--200 & 0.75$\pm$0.04 \\ 
14 Oct 1997 & 15 Oct 1997 & MECS & 14.312 & 1.7--10.5 & 0.122$\pm$0.003
\\
&& PDS & 5.9 & 15--200 & 0.73$\pm$0.11\\
\hline
\end{tabular}
\end{table}

\begin{figure}[h]
\psfig{file=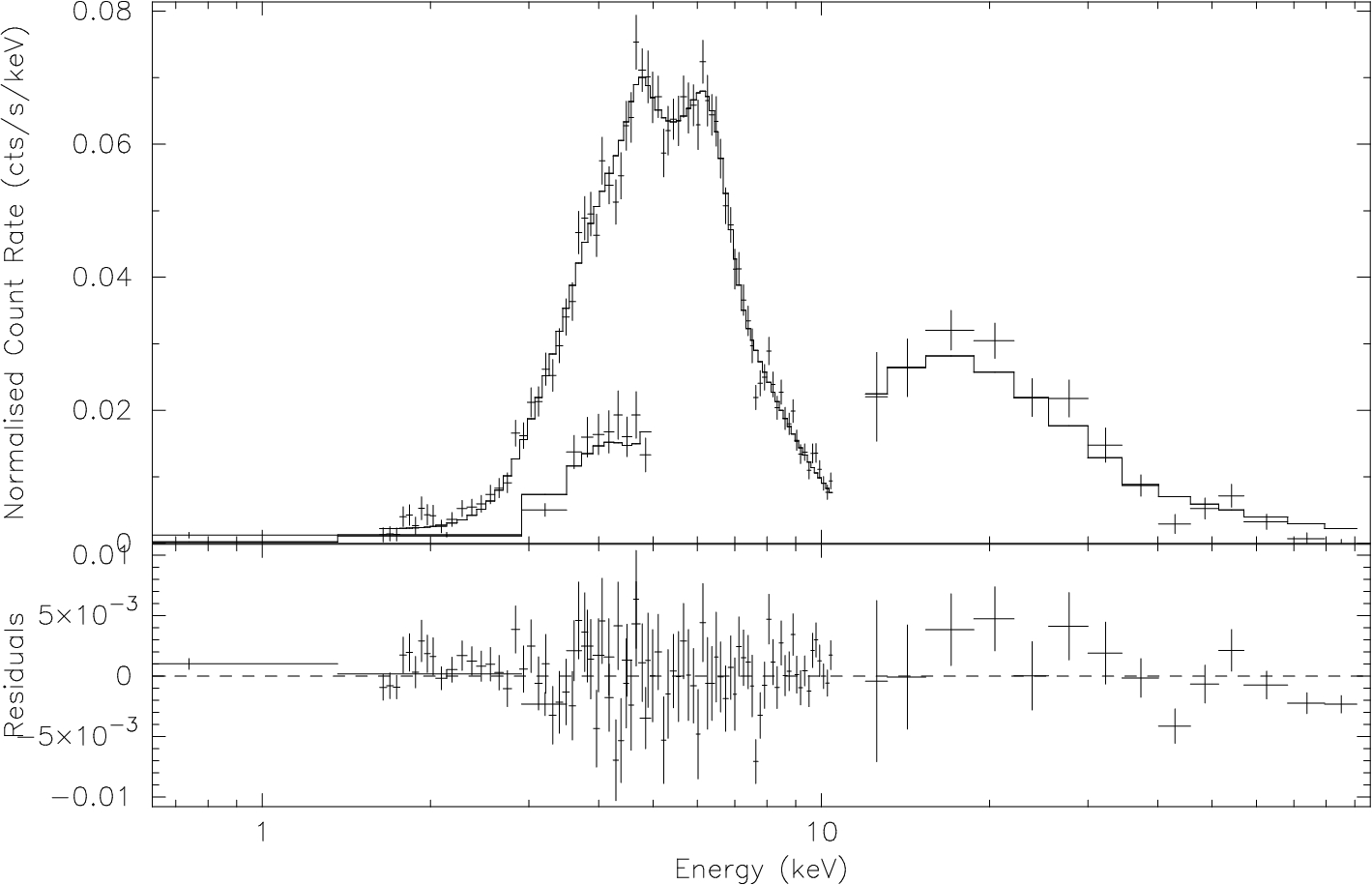,height=6.0truein,angle=-90}
\caption{Best fitting power law plus gaussian emission line 
model for the epoch 1996 BeppoSAX data for ESO103-G35 over the
energy range 0.6--100 keV (Table~\ref{tbres}, Fit 2). The figure shows the data and best fit model 
folded through the instrument response with the residuals in the lower panel.}
\label{figplg}
\end{figure}

\begin{figure}[h]
\psfig{file=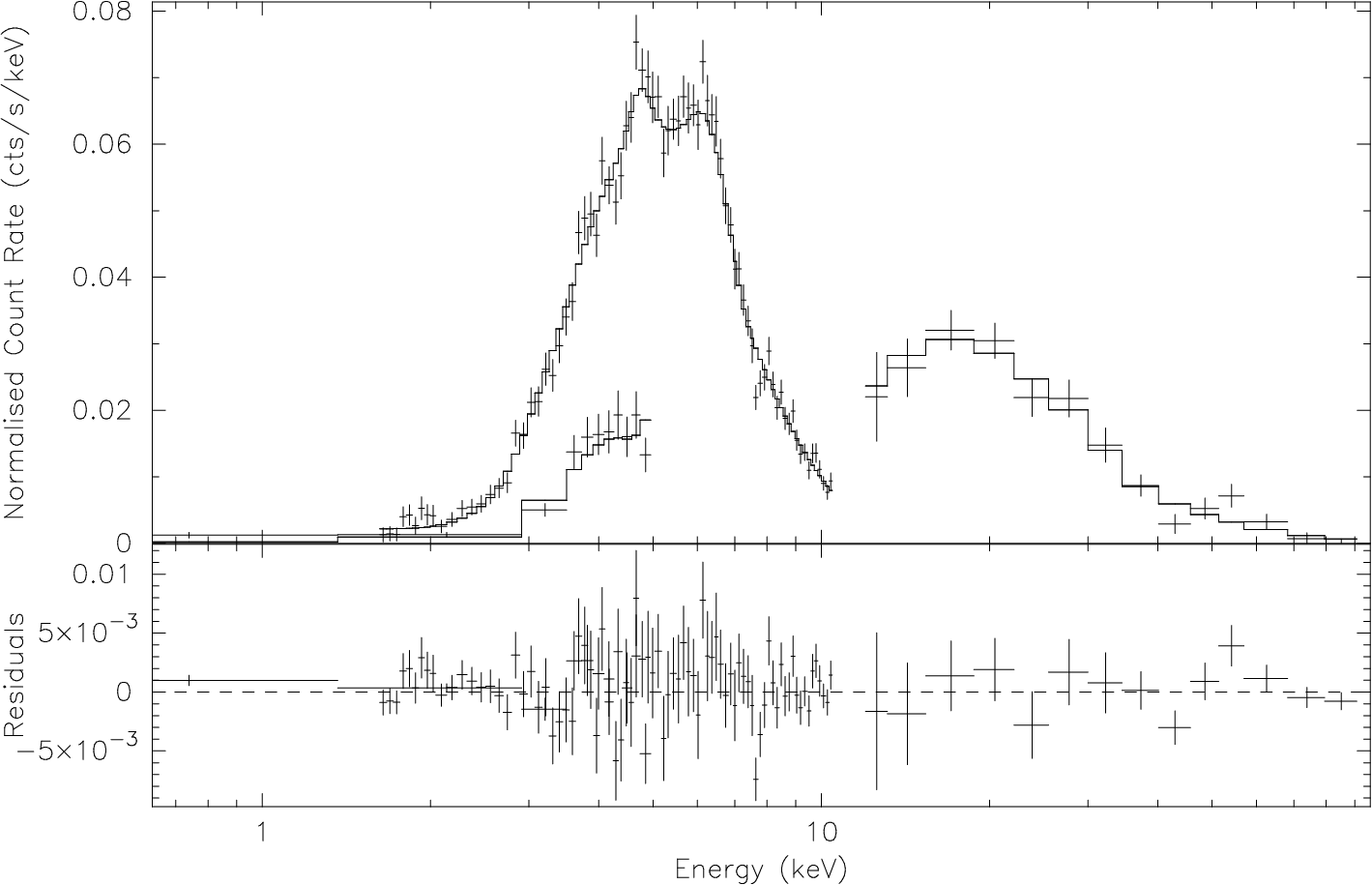,height=6.0truein,angle=-90}
\caption{Best fitting cutoff power law plus gaussian emission line 
for the epoch 1996 BeppoSAX data of ESO103-G35 over the
energy range 0.6--100 keV (Table~\ref{tbres}, fit 3).
The figure shows the data and best fit model 
folded through the instrument response with the residuals in the lower panel.}
\label{figplcut}
\end{figure}

\begin{figure}[h]
\psfig{file=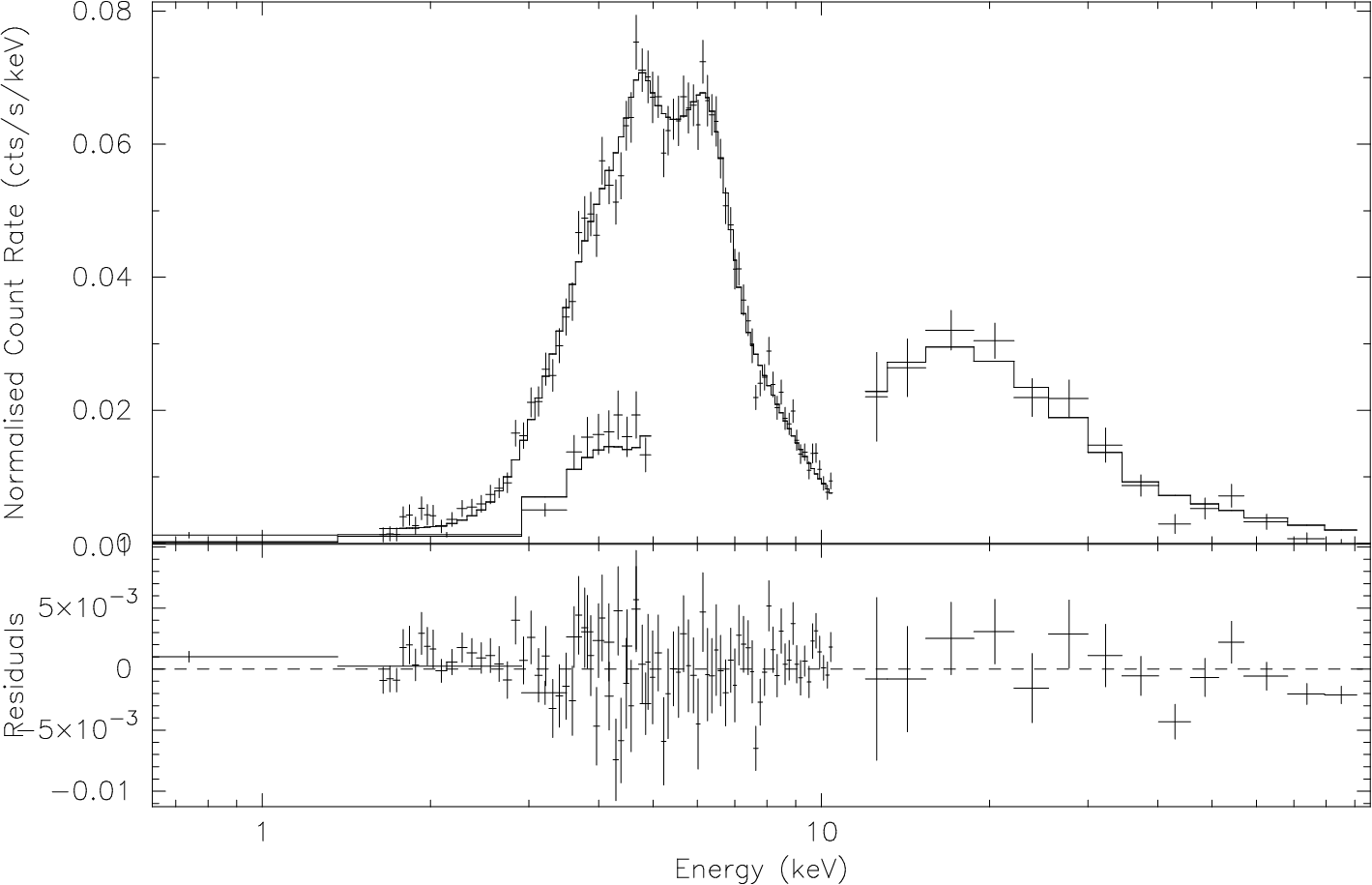,height=6.0truein,angle=-90}
\caption{Best fitting power law plus gaussian emission line plus
reflection model for the epoch 1996 BeppoSAX data of ESO103-G35 over the
energy range 0.6--100 keV (Table~\ref{tbres}, fit 4).
The figure shows the data and best fit model 
folded through the instrument response with the residuals in the lower panel.}
\label{figrefl}
\end{figure}

The spectra were analyzed with the XSPEC spectral fitting package
(Arnaud 1996) using the calibration files provided by the
SDC (March 1998). 
The data from all three instruments were
fitted simultaneously.  Following the recommendation of the SDC, we
constrained
the PDS normalization to
be 0.85 times that of the MECS.
We first fitted the data with an absorbed power-law model (Table~\ref{tbres}, fit 1).
The best fit value of the column density (z=0)
N$_H=(2.07\pm0.007)\times 10^{23}$ cm$^{-2}$ is significantly larger than the Galactic
column density towards the source (N$_H$ \lax $3 \times 10^{20}$ cm$^{-2}$).
The absorbed power law fit showed a clear excess
in both datasets at the position of the Fe K$\alpha$ emission line
($\sim 6.4$ keV). Addition of a Gaussian emission line 
resulted in an acceptable fit (Table~\ref{tbres}, 
fit 2, Figure~\ref{figplg})
with a $\chi^2_\nu$ of 1.16 (239 degrees of freedom, dof). The emission line
equivalent width (EW) is 
=240$^{+80}_{-60}$ eV and it is marginally resolved. 
However the residuals at high energies show an excess suggestive of
curvature. We therefore applied first a high energy cutoff to the powerlaw
(Table~\ref{tbres}, fit 3) and then a reflected component with no high
energy cutoff (Table~\ref{tbres}, fit 4).
The former resulted in a significantly improved $\chi^2_\nu$
(f-test yields 0.01\% significance level). The results are displayed in
Figures~\ref{figplcut} and~\ref{figrefl} respectively.

The 1996 ASCA data (Turner \etal\ 1997, Forster \etal~ 1999) showed an iron edge at about
7.37$^{+0.26}_{-0.22}$ keV with $\tau=0.47^{+0.16}_{-0.12}$, consistent
with earlier observations (Warwick \etal\ 1993).
Since systematic residuals are
present around 7--8 keV in the best fit spectrum (Figure~\ref{figplcut}), we
investigated the presence of excess iron absorption. We added
an edge to the best fit model described above and indeed found an
$>99$\% improvement in the fit (F-test). The best fit edge energy is
7.37$^{+0.15}_{-0.21}$ keV, consistent with ASCA observations in March
1996 (Forster \etal\ 1999), with an
opacity, $\tau=0.24^{+0.06}_{-0.09}$, marginally ($\sim 2 \sigma$)
lower than the ASCA
value. This absorption is likely due to the K-edge of mildly ionized
iron. However, our data quality is not good enough to fit a complex
warm absorber model. We tried using the `absorbi' model in XSPEC, but
could not constrain the parameters.

\begin{table}[h]
\caption{Results of spectral fits to the 1996 dataset$^1$.}
\label{tbres}
\begin{tabular}{lcccc}
\hline
Fits & 1 & 2 & 3 & 4 \\
Model$^2$& PL &PL$+$Gauss & PL$+$HighEcut$+$Gauss & PL+Refl$+$Gauss\\
\hline
\nh\ ($\times 10^{23}$cm$^{-2}$) & 2.07$\pm 0.07$ & 1.92$\pm$0.08 & 
1.79$\pm$0.09 &
1.97$\pm$0.09 \\
F$_{\nu}$(1 keV)$^3$ & 0.024 & 0.019$\pm$0.003  & 0.015$\pm$0.002 &
0.024$\pm 0.005$  \\
\alpe & 0.94$\pm 0.05$ & 0.87$\pm$0.05 & 0.74$\pm0.07$ & 1.00$\pm$0.12 \\
Energy (keV) & - & 6.33$\pm$0.10  & 6.30$\pm 0.10$ & 6.34(fr)  \\
$\sigma$ (keV) & - & 0.32$\pm$ 0.14  & 0.35$\pm$0.14 & 0.30(fr) \\
EW(FeK$\alpha$)(eV) & - & 240$^{+80}_{-60}$ & 290$^{+100}_{-80}$& 
210$^{+60}_{-40}$
\\
E$_{cutoff}$ (keV) & - & - & 29$\pm$10 & - \\
E$_{fold}$ (keV) & - & - & 40$^{+30}_{-20}$ & -  \\
cos $\it i$ (R=1) & - & - & - & 0.2$^{+0.4}_{-0.2}$ \\
$\chi^2_\nu$(dof) & 2.05(103) & 1.51(99) & 1.28(96) & 1.46(101) \\
Flux(2-10 keV)$^4$  & 2.57 & 2.56 & 2.57 & 2.55 \\
\hline
\end{tabular}
\begin{minipage}{6.5in}
1: All errors are quoted at 90\% confidence \\
2: PL: Power Law; Gauss: Gaussian emission line; HighEcut: a high energy
exponential cutoff; Refl: reflection
component \\
3: in photons cm$^{-2}$ s$^{-1}$ keV$^{-1}$ \\
4: in units of $10^{-11}$ erg cm$^{-2}$ s$^{-1}$
\end{minipage}
\end{table}

\begin{figure}[h]
\psfig{file=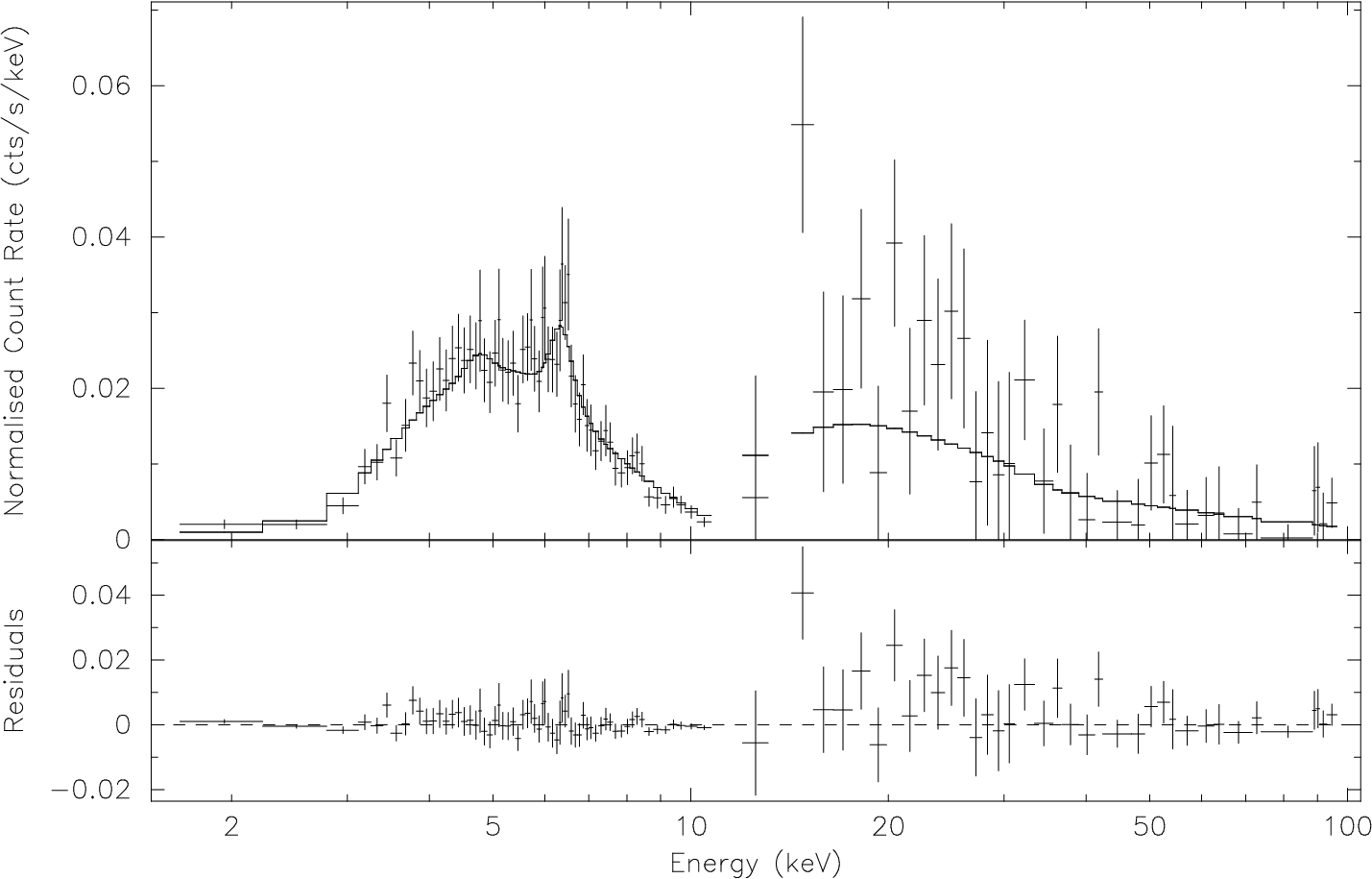,height=6.0truein,angle=-90}
\caption{Best fitting power law plus gaussian emission line 
model for the epoch 1997 BeppoSAX-MECS AND PDS data of ESO103-G35 over the
energy range 0.6--100 keV (Table~\ref{tbres2}, fit 5).
The figure shows the data and best fit model 
folded through the instrument response with the residuals in the lower panel.}
\label{fig97PL}
\end{figure}

The 1997 dataset has a lower MECS count rate and shorter exposure time
so that the data have  much lower signal-to-noise (see Table~\ref{tbobs},
Figure~\ref{fig97PL}).
To quantify the amount of variation, we first fitted a simple
absorbed power law plus Gaussian emission line model 
to the MECS data only. The results are given as fit 5 in Table~\ref{tbres2}
and displayed in Figure~\ref{fig97PL}.
The flux decreased by a factor of
$\sim 2.7$ over the year between the two observations. There is no
evidence for spectral variability or for a variable absorbing column
density. The emission line strength is poorly constrained,
EW=250$^{+150}_{-110}$ eV,
and is consistent with either a constant flux or constant EW over the year.
We therefore are unable to constrain the response time of the
line-emitting material to the continuum variations. There is no
evidence in this spectrum for the line to be broad, though once again the
errors are sufficiently large that it remains consistent with the 1996
line width.

Surprisingly,  the PDS data in the 1997 dataset show a count rate similar to
the 1996 value (Table~\ref{tbobs}). We fitted the combined MECS+PDS
data for 1997. An absorbed  power law plus Gaussian line
fit to the combined  dataset shows a flatter slope
(see Table~\ref{tbres2}, fit 6) and gives an acceptable $\chi^2$ value.
The resulting spectral index (\alpe$\sim$0.5) is much flatter than in 1996
and would imply a significant hardening of the spectrum as the flux decreases.
Since there is no other evidence for spectral variability in ESO103-G35, we
seek an alternative interpretation.

Partial covering of a hard continuum source can be ruled out by the
lack of variability in the PDS data. However, since the PDS
data would be dominated by any reflected component, an alternative
interpretation of this pattern of variability is that of a reflected component
which remains constant while the power law component varies. Fit 7
(Table~\ref{tbres2}) shows the best fitting power law plus reflection
plus Gaussian line model to the MECS+PDS 1997 dataset. This gives
a marginally better fit but the data are not of sufficiently high
quality to provide strong contraints on the parameters.
For this to be realistic, the reflecting
medium must respond to a change in incident continuum
with some delay or be seeing a constant continuum which 
is invisible to us. This latter could
occur, for example, if the flux variation were due to absorption by
material between us and the reflecting material. However the spectrum shows
no evidence for a variation in the column density to support this scenario.
The most likely explanation is that of a reflected component,
undetectable in the earlier dataset and with a
delayed response to continuum variations. With only two data points,
we cannot be sure whether further continuum 
variations occurred during the period in between our observations
to which the reflected component is responding. Thus we cannot place
any meaningful constraints on the distance of the reflecting region from
the continuum source.

\begin{table}[h]
\caption{Results of spectral fits to the 1997 dataset$^1$.}
\label{tbres2}
\begin{tabular}{lccc}
\hline
Fits & 5 & 6 & 7\\
Model$^2$ & PL$+$Gauss$^3$ &PL$+$Gauss & PL+Refl$+$Gauss  \\
\hline
\nh\ ($\times 10^{23}$cm$^{-2}$) & 2.02$\pm$0.28 & 1.68$^{+0.23}_{-0.21}$ &
1.78 \\
F$_{\nu}$(1 keV)$^4$ & 0.008$^{+0.0078}_{-0.0034}$ & 0.003$_{-0.001}^{+0.002}$
& 0.006 \\
\alpe & 0.9$\pm$0.3 & 0.49$^{+0.19}_{-0.17}$ & 0.93\\
Energy (keV) & 6.45$_{-0.19}^{+0.11}$ & 6.43$_{-0.25}^{+0.11}$ & 6.42 \\
$\sigma$ (keV) & $<0.3$  & $<0.4$ & $<0.4$ \\
EW(FeK$\alpha$)(eV) & 250$^{+150}_{-110}$ & 260$^{+180}_{-110}$ & 193 \\
R (cos i=0.45) & & & 4$^{+7}_{-3}$ \\ 
$\chi^2_\nu$(dof) & 0.66(65) & 0.87(96) & 0.78(94)\\
Flux(2-10 keV)$^5$  & 0.95 & 0.98 & 0.97 \\
\hline
\end{tabular}
\begin{minipage}{6.5in}
1: All errors are quoted at 90\% confidence \\
2: PL: Power Law; Gauss: Gaussian emission line; Refl: reflection
component \\
3: MECS data only \\
4: in photons cm$^{-2}$ s$^{-1}$ keV$^{-1}$ \\
5: in units of $10^{-11}$ erg cm$^{-2}$ s$^{-1}$
\end{minipage}
\end{table}

\section{Discussion}
\subsection{Absorption}
The BeppoSAX observations confirm the presence of a large column density of neutral 
absorbing material (\nh = $2.0\pm 0.1\times 10^{23}$ cm$^{-2}$) in the 
X-ray spectrum of ESO103-G35. The column density measured here is consistent
with the higher column density reported in the EXOSAT data (Warwick \etal\ 
1988)
as well as that from more recent ASCA data (Forster \etal\ 1999,
Turner \etal\ 1997).
There is no evidence for variation in this absorbing column between the
1996 and 1997 observations. Figure~\ref{fgflux} and Table 4 show
a compilation of \nh\
measurements over a 13 year period (1984$-$1997).
They are generally consistent
with one another, the largest deviation\footnote{apart from the rapid change
reported by Warwick \etal\ (1988) which may have been due to a spurious
source in the {\it Ginga} data ($\S 1$)} is $\sim 2 \sigma$ in 1991.

Absorbing material with column densities $\sim 10^{23}$ cm$^{-2}$
is typically seen in the subset of Seyfert 2 
galaxies which show evidence for a hidden Seyfert 1 (Turner \etal\ 1997,
Awaki 1997).
It is consistent with a line-of-sight passing through an edge-on, optically
thick accretion
disk/torus of neutral material such as is commonly believed to be
present in Seyfert galaxies (Antonucci \& Miller 1985). The presence of
water maser emission in ESO103-G35 (Braatz \etal\ 1997) indicates that
our line-of-sight intersects the accretion disk/torus accurately edge-on.
In the absence of warps in the disk/torus, the X-ray absorption 
measures its full column density.

We also detect excess absorption due to ionized iron,
most likely due to FeII--FeXV,  consistent with earlier Ginga and ASCA results.
The observed opacity $\tau=0.24^{+0.06}_{-0.09}$ is somewhat lower than the
ASCA value (Turner \etal\ 1997, Forster \etal\ 1999). While our data
quality are insufficient to constrain a full warm absorber model,
the presence of the edge requires ionized material along our line-of-sight.
Given the
accurately edge-on geometry, the ionized material most likely lies
within the inner regions of the accretion disk/torus close to the
central continuum source.
One possible alternative would be a warp in the disk/torus
so that our line-of-sight does not pass through the full geometric size of the
disk. A sufficiently strong warp would allow the possibility of a
separate warm absorber above or below the accretion disk/torus at a
position closer to the central source than the intersection of the
warped disk with our line-of-sight.

\begin{table}
\caption{Compilation of spectral parameters for earlier
observations of ESO103-G35}
\label{his:tab}
\begin{tabular}{llllllll}
\hline
Satellite & Instrument &Date & Flux (2-10 keV)$^4$ & \nh\ & \alpe &
Ref$^1$.& \% Error  \\
\hline
EXOSAT & ME & 247/1983 & 1.81$\pm$0.10& 2.28$^{+1.37}_{-.98}$&0.90$^{+0.96}_{-0
.34}$& 1
& 90\\
EXOSAT & ME & 110/1984 & 2.18$\pm$0.07& 2.47$^{+.81}_{-.65}$ 
&1.54$^{+0.71}_{-0.56}$& 1
& 90\\
EXOSAT & ME & 124/1985 & 2.46$\pm$0.09& 2.37$^{+.90}_{-.72}$ 
&1.23$^{+0.78}_{-0.60}$& 1
& 90\\
EXOSAT & ME & 214/1985 & 2.68$\pm$0.18& 0.71$^{+.73}_{-.42}$ 
&0.24$^{+0.76}_{-0.31}$&
1 & 90 \\
EXOSAT & ME & 225/1985 & 1.90$\pm$0.11& 0.81$^{+.54}_{-.39}$ 
&0.57$^{+0.76}_{-0.44}$& 1
& 90\\
EXOSAT & ME & 247/1985 & 1.22$\pm$0.10& 1.67$^{+1.95}_{-1.01}$ 
&1.08$^{+1.82}_{-0.54}$& 1
& 90 \\
Ginga & LAC & 268/88 & 2.1$\pm 0.7 $ & 
$1.76^{+0.41}_{-0.31}$
&0.76$^{+0.19}_{-0.17}$ & 2 & 90$^2$\\
Ginga & LAC & 102/91 & 1.9$\pm0.3$ & 
$3.55^{+0.72}_{-0.56}$
&0.84$^{+0.08}_{-0.11}$ & 3 & 90$^3$\\
ASCA & SIS+GIS & 246/94 & 0.9$\pm 0.4 $ &
1.565$^{+0.288}_{-0.173}$ & 0.37$^{+0.33}_{-0.21}$& 4 & 90$^3$ \\ 
ASCA & SIS+GIS & 246/94 & 1.42$\pm0.05^6$ & 2.16$^{+0.40}_{-0.35}$ &0.89$^{+0.35}_{-0.40}$ &
5 & 90 \\
ASCA & SIS+GIS & 269-270/95 & 2.36$\pm0.13^6$ & 1.68$^{+0.54}_{-0.48}$ &0.31$^{+0.59}_{-0.57}$ &
5 & 90 \\
ASCA & SIS+GIS & 078/96 & 2.38$\pm0.06^6$ & 2.16$^{+0.26}_{-0.25}$ &1.08$^{+0.29}_{-0.28}$ &
5 & 90 \\
\hline
\end{tabular}
\begin{minipage}{6.5in}
1: References: 1: Warwick \etal\ 1988, 2: Warwick \etal\ 1993, 3: Smith
\& Done 1996, 4: Turner \etal\ 1997, 5: Forster \etal\ 1999\\
2: Average of quoted error on normalization \\
3: Based upon quoted error in slope \\
4: In units of $10^{-11}$ erg cm$^{-2}$ s$^{-1}$\\
5: In units of $10^{23}$ cm$^{-2}$ \\
6: Errors from photon statistics only
\end{minipage}
\end{table}

\subsection{Spectral Shape}
\begin{figure}[h]
\psfig{file=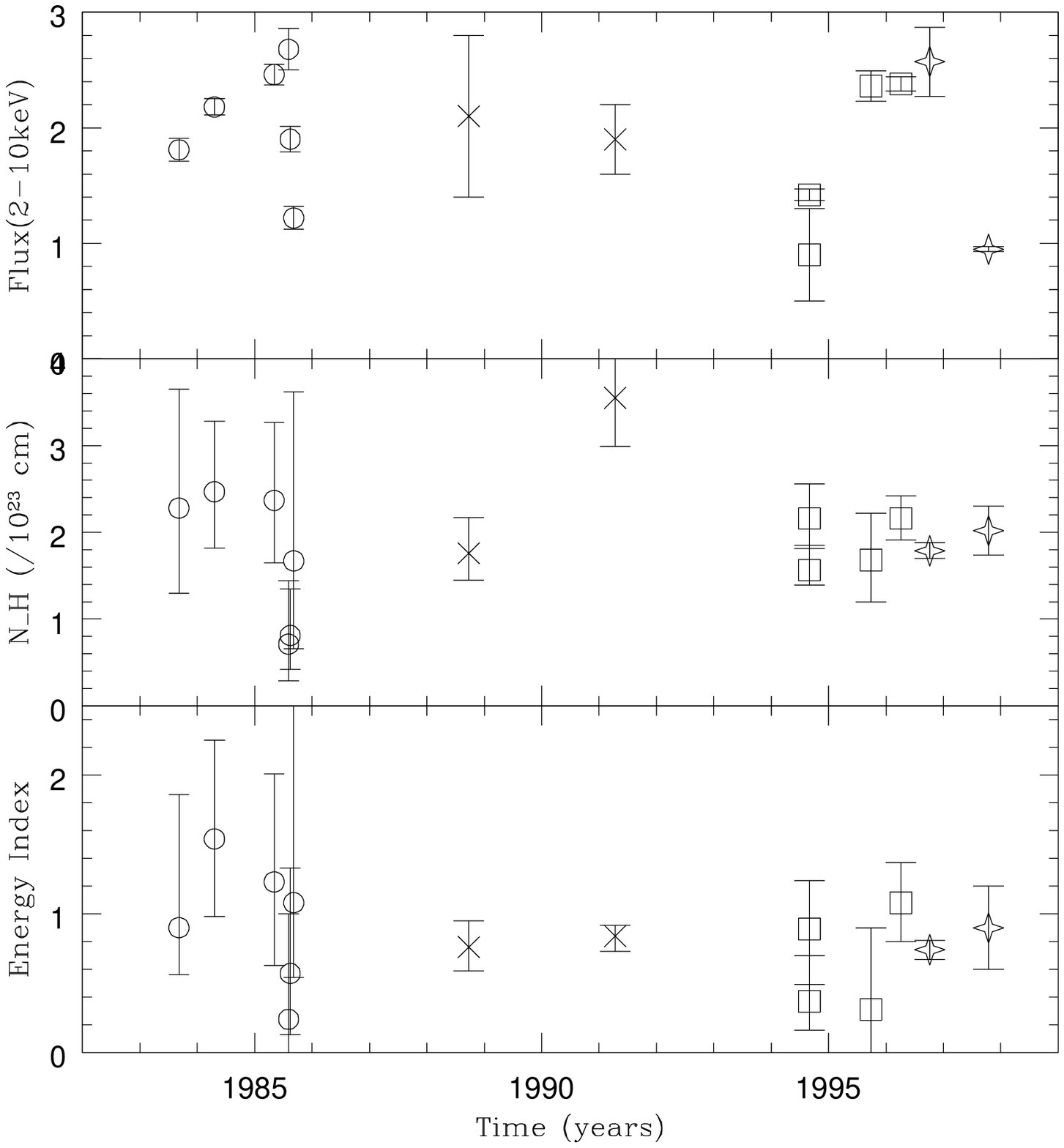,height=6.0truein}
\caption{Time history of the broad-band (2--10 keV) flux
(in units of $10^{-11}$ erg cm$^{-2}$ s$^{-1}$), absorbing column 
density
and energy spectral index of ESO103-G35 from 1983$-$1997.
Significant flux variability is present on timescales of a few months
while the spectral index remains constant within the
errors. The absorbing column has varied on short timescales in the past but
in general remains fairly constant as well. The different datasets are
indicated as follows: circle: EXOSAT, cross: Ginga, square: ASCA,
star: BeppoSAX.  
}
\label{fgflux}
\end{figure}

The best fitting model is a cutoff power law with slope, $\alpha_E
= 0.74\pm$0.07 (Table~\ref{tbres}, fit 3),
a cutoff with energy of 29$\pm$10 keV
and e-folding energy 40$^{+30}_{-20}$keV.
This model shows a resolved Fe K$\alpha$ emission line at 6.3$\pm$0.1 keV
and equivalent width 290$^{+100}_{-80}$ eV.
Figure~\ref{fgflux} shows the lack of significant
variation in spectral parameters
over a 13 year period.

In the hard X-ray, where the effects of the large absorbing column are
negligible, the scenario of Seyfert 2 galaxies being edge-on Seyfert 1s
predicts that both classes of galaxy look similar.
Hard X-ray and $\gamma$-ray data for radio-quiet Seyfert 1 galaxies 
typically show a power law spectrum with spectral index, $\alpha_E \sim 0.9$,
similar to that reported here for ESO103-G35 but
extending to energies $> 100$ keV (Gondek \etal\ 1996) rather than 
$\sim 30$ keV. This hard X-ray spectrum can be
explained by emission from an optically thin, relativistic, thermal plasma
in a disk corona or by a non-thermal plasma with a power-law injection of 
relativistic electrons (Gondek \etal\ 1996). 
The lower high energy cutoff suggests this component has 
a higher optical depth in ESO103-G35. 
In addition the X-ray spectra of radio-quiet Seyfert 1 galaxies
typically include a contribution from Compton reflection in warm material
which is superposed on the underlying power law in the energy range
10--100 keV (Gondek \etal\ 1996). The 1996 observation favors
a cutoff power law over a reflection model suggesting that this latter
component is, at most, weak in ESO103-G35 in its brighter state. However
the relatively strong hard flux in the otherwise lower luminosity 1997 dataset
suggests the presence of a reflection component in 1997.
Since the reflection
model includes a large number of parameters, it is not possible to
constrain them usefully. However if we assume an isotropic source
situated above the accretion disk, no high energy cutoff and solar abundances
we deduce an inclination angle for the disk in the range $50^{\circ} -90^{\circ}$,
consistent with ESO103-G35 being a Sy1.9/2 in unified models and with
the accurately edge-on geometry implied by the presence of water maser
emission. 

\subsection{Luminosity Variation}


Figure~\ref{fgflux} shows the historical light curve of ESO103-G35
since 1983 including the current BeppoSAX observations (1997).  It is clear
that flux variations of factors of a few are common on timescales of
months. The variation seen by ASCA and BeppoSAX between 1994 and 1997 shows
a slow rise and rapid fall quite similar to that seen by {\it Ginga} between
1983 and 1985. The flux level
in the 1996 observation (F(2--10 keV) = 2.57$\pm$0.07$\times 10^{-11}$ erg
cm$^{-2}$ s$^{-1}$) indicates a luminosity, L$_x \sim 2 \times 10^{43}$ 
erg s$^{-1}$ (H$_0$=50, q$_0$=0).

\subsection{Fe K$\alpha$ Emission Line}
The emission line is cold (6.3$\pm$0.1 keV), but consistent with being
mildly warm, and marginally resolved (Table~\ref{tbres}).  Adopting an
MECS instrument energy resolution of ($\sigma$) 0.21 keV at 6.3 keV
(SAX Observers' Handbook), the intrinsic width of the line is
estimated to be 0.28$\pm$0.12 keV which yields a FWHM of $\sim (31 \pm
12) \times 10^{3}$ km s$^{-1}$ for the Fe K$\alpha$ line.  This
indicates an origin in neutral, high-velocity material.
The EW, 290$^{+100}_{-80}$ eV,
is within the wide range of observed values for Seyfert 2 galaxies
and is consistent with fluorescence in the cold absorber, the accretion
disk/torus,
which is attenuating the observed X-ray continuum (Awaki \etal\ 1991, Turner
\etal\ 1997).
%
%


The emission line in the second dataset is not well-constrained. It is
consistent with constant flux or with constant equivalent width over
the one year period so we cannot determine if any response to the
continuum variation has occurred. 
From the
$\sigma=0.32$ keV (Table~\ref{tbres}), the FWHM of the line is about 30,000 km
s$^{-1}$. 

If the line width is dynamic, indicating a virial velocity around
the 10$^7$ M$_{\odot}$
black hole, then the radial distance to the emitter is about 0.01 pc.
We note that an alternative interpretation of the broad Fe K-$\alpha$ line
is that of a mix of neutral and ionized material. This would be consistent
with the earlier suggestion (\S 3.1)
of ionized material in the inner regions of the accretion disk/torus being
responsible for the ionized Fe-K edge. 
Fits to the 1994 ASCA dataset
with three narrow lines covering a range of ionization
were statistically indistinguishable from
those with a broadened, neutral line (Turner \etal\ 1997). 

\subsection{The High Energy Cutoff}
BeppoSAX, with its high energy response, has provided information on the
high energy spectrum of a number of AGN for the first time. 
The dispersion and errors in the observed values of high energy cutoff
of AGN are high but values typically range from
100$-$400 keV (Matt 1998). The high energy spectrum of AGN is a critical
parameter for matching both the turnover at $\sim 30$ keV and the
high energy source counts when modelling the CXRB.  Successful models
generally adopt a value of 300 keV and require strong evolution in the
number of absorbed AGN with redshift (Gilli, Risaliti \& Salvati 1999).
However, the high energy cutoff determined from our
BeppoSAX data for ESO103-G35 (29$\pm$10 keV)
is significantly lower, calling into question the validity of the modelling
to date. More realistic modelling of the
distribution of AGN spectral
properties, including the high energy cutoff, is required to make an
accurate assessment of the AGN contribution to the CXRB (Yaqoob 2000).

\section{Conclusions}

ESO103-G35 is an X-ray bright, Sy 1.9/2.0 galaxy containing 
one of strongest water maser sources known implying that it is viewed
accurately edge-on. The presence of cold molecular material along the line
of sight to the active nucleus supports this picture. The amount of
X-ray absorption has been $\sim$constant
over the past 15 years and, at $2\times 10^{23}$ cm$^{-2}$, it is typical
of Seyfert 2 galaxies and consistent with a line-of-sight passing
through the edge-on accretion disk/torus also responsible for the maser
emission. 

We observe a marginally resolved Fe-K$\alpha$ emission
line whose width, if dynamic, implies a radial distance to the absorber of
about 0.01 pc, similar to that expected for an obscuring torus.
Alternatively, the width may be due to a range of ionization states
in the emitting material consistent with the ionized Fe K edge at
7.37$^{+0.15}_{-0.21}$ keV.
The Fe K$\alpha$ line strength (EW=290$^{+100}_{-80}$ eV) and width
are consistent with origin in an accretion disk/torus
through which our line-of-sight is passing.
The presence of both cold and warm (ionized)
absorbing (and emitting) material suggests that the inner
parts of the torus/disk are mildly ionized
while the outer parts are cold. 

The flux level in the energy range 0.6$-$10 keV decreased by a factor
$\sim 2.7$ between October 1996 and October 1997 with no evidence for a
change in spectral shape or absorption. However the 10-50 keV flux did
not change between the two observations suggesting either a hardening of
the hard continuum or the presence of a reflection component
(undetectable at the higher flux level) which 
reacts to continuum changes with some delay. Given the paucity of
datapoints, we are unable to place meaningful constraints on the
distance of the reflecting material from the continuum source. 

The 1996 dataset shows a high energy cutoff at $29\pm 10$ keV. This is
significantly lower than the values $\sim 300$ keV believed typical to
date (Gondek \etal\ 1996, Matt 1998). The high energy spectrum of AGN is
critical for modelling the AGN contribution to the CXRB so that this
result emphasizes the need to include
a realistic distribution of cutoff energy in these models.
\noindent
\acknowledgments
BJW and SM gratefully acknowledge the financial support of
NASA grants:  NAG5-7064 (BeppoSAX), NAG5-3249 (LTSA).

\end{document}